\documentclass[dvips]{acta}
\usepackage{supertabular,lscape,epsfig}
\usepackage{amssymb}
\usepackage{amsmath}
\SetPages{1}{11}
\SetVol{63}{2013}

\begin{document}

\begin{Titlepage}

\Title { Fifty Years of TT Arietis }

\Author {J.~~S m a k}
{N. Copernicus Astronomical Center, Polish Academy of Sciences,\\
Bartycka 18, 00-716 Warsaw, Poland\\
e-mail: jis@camk.edu.pl }

\Received{  }

\end{Titlepage}

\Abstract { Results of photometric observations of the permanent negative 
superhumper TT Ari in 1961/62 and 1966 are presented. 
Together with data from the literature they are used to discuss 
the negative superhump amplitudes $A_{nSH}$ and the amplitudes $A_{beat}$ of 
the modulation with the beat period $P_{beat}$. 
Both amplitudes are shown to vary considerably from one season to another. 
Three correlations are found: (1) between $A_{nSH}$ and $A_{beat}$, 
(2) between $A_{nSH}$ and $P_{nSH}$, and (3) between $A_{beat}$ and $P_{beat}$.  
}
{binaries: cataclysmic variables, stars: individual: TT Ari }

\section { Introduction } 

The variability of TT Ari = BD+14{$^\circ$}341 was discovered by Strohmeier, 
Kippenhahn and Geyer (1957). In the fall of 1961 Herbig (1961) took several 
spectrograms of the star and found that its spectrum consists of 
a hot continuum and weak emission lines of hydrogen; this suggested that 
BD+14{$^\circ$}341 could be a nova-like object. 
Following Dr. Herbig's suggestion the present author, then at the Lick Observatory, 
observed the star photometrically in December 1961 and January 1962  
and found that on shorter time scales its variability consists of three components: 
(1) Periodic variations with $P=0.1329$d, or $\sim 3^h$12$^m$, and full 
amplitude of about $2A\approx 0.2$mag, often referred to as "3-hour" variations. 
(2) Transient, quasi-periodic fluctuations with periods 
between 14 and 20 minutes and full amplitudes up to $2A\approx 0.2$mag. 
(3) Rapid flickering with amplitudes up to 0.1 mag and time scales of the order of 
1 min. Those findings were later confirmed by results of two other series of photometric 
observations: in 1966 by the present author at the Observatoire de Haute Provence (OHP) 
and in 1967 by Dr. K. St{\c e}pie{\'n} at the Lick Observatory.  
Only preliminary results of those three series of photometric observations were 
published by Smak and St{\c e}pie{\'n} (1969, 1975). 

It can be added that the "3-hour" variations of TT Ari were in fact the first 
superhumps ({\it negative} superhumps in this case) ever observed although they were 
identified as such only 30 years later (cf. Patterson et al. 1993). 

In the following years TT Ari was observed photometrically and spectroscopically 
by many authors. 
The first extensive spectroscopic investigations of TT Ari by Cowley et al. (1975) 
revealed that its orbital period is $P_{orb}=0.1375$d, i.e. about 3 percent longer 
than the photometric period. This was confirmed later by Thorstensen, Smak and 
Robinson (1985) and by Wu et al. (2002) who determined 
$P_{orb}=0.13755040\pm 0.00000017$d. 

Semeniuk et al. (1987) found that the mean brightness of TT Ari observed in 1966  
(see Section 3) varied also with the "4-day" beat period resulting from 
the combination of the orbital and negative superhump periods. 
So far, however, the existence of this "4-day" modulation was confimed only 
by Udalski (1987) and Kraicheva et al. (1997), but not by other observers. 

In 1997 an unexpected transition occured from negative superhumps to common superhumps 
with $P_{SH}=0.1492$d, about 8 percent longer than the orbital period 
(Kraicheva et al. 1999, Skillman et al. 1998). 
The common superhumps disappeared and the negative superhumps begun to reappear 
again in 2005 (Andronov et al. 2005, Kim et al. 2009) and in 2007 they were observed 
with the MOST satellite (Vogt et al. 2013). 

TT Ari is also a member of the VY Scl subtype of CV's showing the so-called 
low states, extending over months or years, during which it declines in brightness 
from $V\approx 10.6$ in its high state down to $V\sim 17$ (cf. Hudec, Huth, and 
Fuhrmann 1984, Shafter et al. 1985, Wenzel et al. 1992). 
The two most recent low states occured in 1980-1984 and in 2009/2010. 
It can be added that no superhumps are observed during those low states. 

The purpose of the present paper is twofold: 
(1) to present in more detail the results of the 1961/62 Lick and 1966 OHP 
photometry (Sections 2 and 3), and (2) to re-analyze the available photometric data  
in order to clarify the problem of the "4-day" variations (Sections 4 and 5).

\section { The 1961/62 Lick Light Curves } 

Observations were made with the Crossley Reflector and standard (at that time!) 
photometric equipment. The comparison star was BD+14{$^\circ$}336 with 
V$=8.944\pm 0.009$, B-V$=+0.220\pm 0.003$, U-B$=+0.108\pm 0.004$ 
(based on 18 measurements during 8 nights). 
On one night TT Ari was observed in three colors allowing the determination 
of its mean magnitude and colors: $<V>\approx 10.6$, $<(B-V)>\approx -0.05$, 
$<(U-B)>\approx -0.95$. On the remaining nights it was observed in one color, 
either in yellow or in ultraviolet. Altogether 1570 data points were obtained. 
Results were expressed in the instrumental system in the form 
$\Delta m=$TT Ari--BD+14{$^\circ$}336. 

Fig.1 shows, as an example, the ultraviolet light curve observed on January 15 UT, 
1962. This is the light curve which was analyzed by Williams (1966) who found  
no periodicities in its first part and three periodic components 
with periods 13.9, 17.6, and 42.2 min. during the second part. No oscillations 
with $P=27$ min. suggested by Semeniuk et al. (1987) were present.  

\begin{figure}[htb]
\epsfysize=15.0cm 
\hspace{0.5cm}
\epsfbox{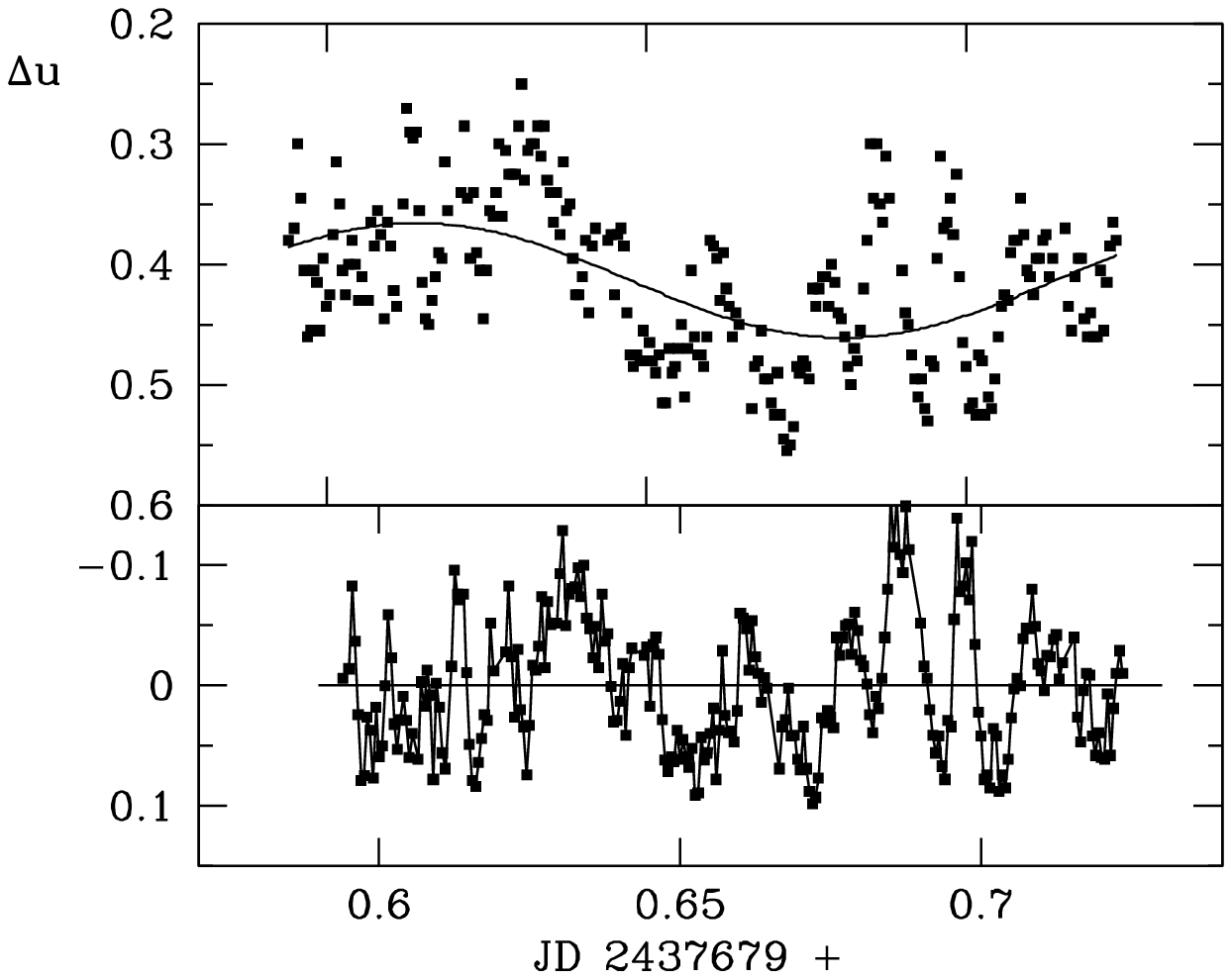} 
\vskip -7truecm
\FigCap { {\it Top:} Ultraviolet light curve of TT Ari on January 15 UT, 1962. 
Solid line is the best fit cosine curve. {\it Bottom:} The residuals after 
subtracting the cosine curve. }
\end{figure}

The parameters describing the negative superhumps, i.e. the moments of maxima 
and minima and the corresponding magnitudes, were determined by fitting 
the cosine curves to the points observed on a given night. 
Results are listed in Table 1. 
Using those moments of maxima nad minima we find the following elements

\beq 
{\rm Maximum}~=~{\rm JDhel.}2437646.6514(19)~+~0.132896(12)\times E~.
\eeq

\begin{table}[h!]
{\parskip=0truept
\baselineskip=0pt {
\medskip
\centerline{Table 1}
\medskip
\centerline{ Maxima and Minima of TT Ari in 1961/62 and 1966 }
\medskip
$$\offinterlineskip \tabskip=0pt
\vbox {\halign {\strut
\vrule width 0.5truemm #&	
\enskip#\enskip&	        
\vrule#&			
\enskip#\enskip&                
\vrule#&			
\enskip#\enskip&                
\vrule width 0.5truemm #&       
\enskip#\enskip&                
\vrule#&			
\enskip#\enskip&                
\vrule#&			
\enskip#\enskip&                
\vrule width 0.5 truemm # \cr	
\noalign {\hrule height 0.5truemm}
&&&&&&&&&&&&\cr
&\hfil Maxima\hfil&&C&&\hfil$\Delta$m\hfil&&\hfil Minima\hfil&&C&&\hfil$\Delta$m\hfil&\cr
&&&&&&&&&&&&\cr
\noalign {\hrule height 0.5truemm}
&&&&&&&&&&&&\cr
&2437000+ \hfil	   && && 	      &&2437000+ \hfil    && &&	             &\cr
&646.6547$\pm$.0018&&V&&1.537$\pm$.011&&655.6129$\pm$.0043&&V&&1.808$\pm$.015&\cr
&655.6863$\pm$.0020&&V&&1.640$\pm$.008&&655.7527$\pm$.0016&&V&&1.812$\pm$.009&\cr
&656.7563$\pm$.0029&&V&&1.714$\pm$.009&&656.6851$\pm$.0016&&V&&1.846$\pm$.007&\cr
&660.7362$\pm$.0020&&U&&0.397$\pm$.010&&660.6723$\pm$.0015&&V&&1.858$\pm$.008&\cr
&664.7239$\pm$.0016&&U&&0.364$\pm$.010&&660.6720$\pm$.0015&&B&&1.594$\pm$.008&\cr
&675.6269$\pm$.0015&&U&&0.350$\pm$.010&&660.6744$\pm$.0013&&U&&0.593$\pm$.007&\cr
&679.6156$\pm$.0033&&U&&0.363$\pm$.009&&664.6569$\pm$.0018&&U&&0.557$\pm$.010&\cr
&692.6255$\pm$.0023&&U&&0.451$\pm$.014&&672.6350$\pm$.0027&&U&&0.552$\pm$.023&\cr
&2439000+\hfil     && &&              &&679.6790$\pm$.0024&&U&&0.463$\pm$.008&\cr
&360.6245$\pm$.0023&&U&&0.372$\pm$.009&&2439000+ \hfil    && &&		     &\cr
&375.6185$\pm$.0015&&U&&0.370$\pm$.010&&375.5536$\pm$.0014&&U&&0.584$\pm$.009&\cr
&376.5492$\pm$.0010&&U&&0.420$\pm$.010&&376.6138$\pm$.0009&&U&&0.700$\pm$.008&\cr
&377.6179$\pm$.0016&&U&&0.430$\pm$.012&&377.5460$\pm$.0012&&U&&0.749$\pm$.011&\cr
&378.5445$\pm$.0013&&U&&0.364$\pm$.008&&378.6102$\pm$.0018&&U&&0.548$\pm$.009&\cr
&&&&&&&&&&&&\cr
\noalign {\hrule height 0.5truemm}
}}$$
}}
\end{table}

The mean amplitudes of the negative superhumps, $A_{nSH}$, and of the "4-day" 
modulation with the beat period, $A_{beat}$, were determined directly by fitting 
the following formula to all data points

\beq
\Delta m~=~<\Delta m>~-~A_{nSH}~\cos \phi_{nSH}~
      -~A_{beat}~\cos (\phi_{beat}-\phi_{beat}^{max})~.  
\eeq 

\noindent 
This formula requires several comments. 
(1) The parameters $<\Delta m>=<\Delta y>$ or $<\Delta u>$, and 
$A_{nSH}=A_{nSH}^V~{\rm or}~A_{nSH}^U$ were determined independently for the two colors.  
(2) The amplitude of the "4-day" modulation $A_{beat}$ was assumed to be identical 
in V and U, this assumption being based on the commonly adopted interpretation 
of negative superhumps (see Section 6). 
(3) The beat phase $\phi_{beat}$ was calculated using the beat period $P_{beat}$ 
related to the orbital and negative superhump periods: 

\beq
1/P_{beat}~=~1/P_{nSH}~-~1/P_{orb}~. 
\eeq

The results of this "global" fit are: $<\Delta y>=1.716\pm 0.003$ or $<V>=10.66$, 
$A_{nSH}^V=0.086\pm 0.004$, $<\Delta u>=0.422\pm 0.003$, $A_{nSH}^U=0.074\pm 0.003$, 
and $A_{beat}=0.067\pm 0.003$. 
To avoid possible confusion it should be added that the "full amplitudes" of those 
variations are obviously 2 times larger than the amplitudes defined by Eq.(2).  

Fig.2 shows the maximum and minimum magnitudes from Table 1 plotted as a function 
of the beat phase. Shown also are the best fit cosine curves obtained from the 
"global" fit (Eq.2). 

\begin{figure}[htb]
\epsfysize=12.0cm 
\hspace{2.0cm}
\epsfbox{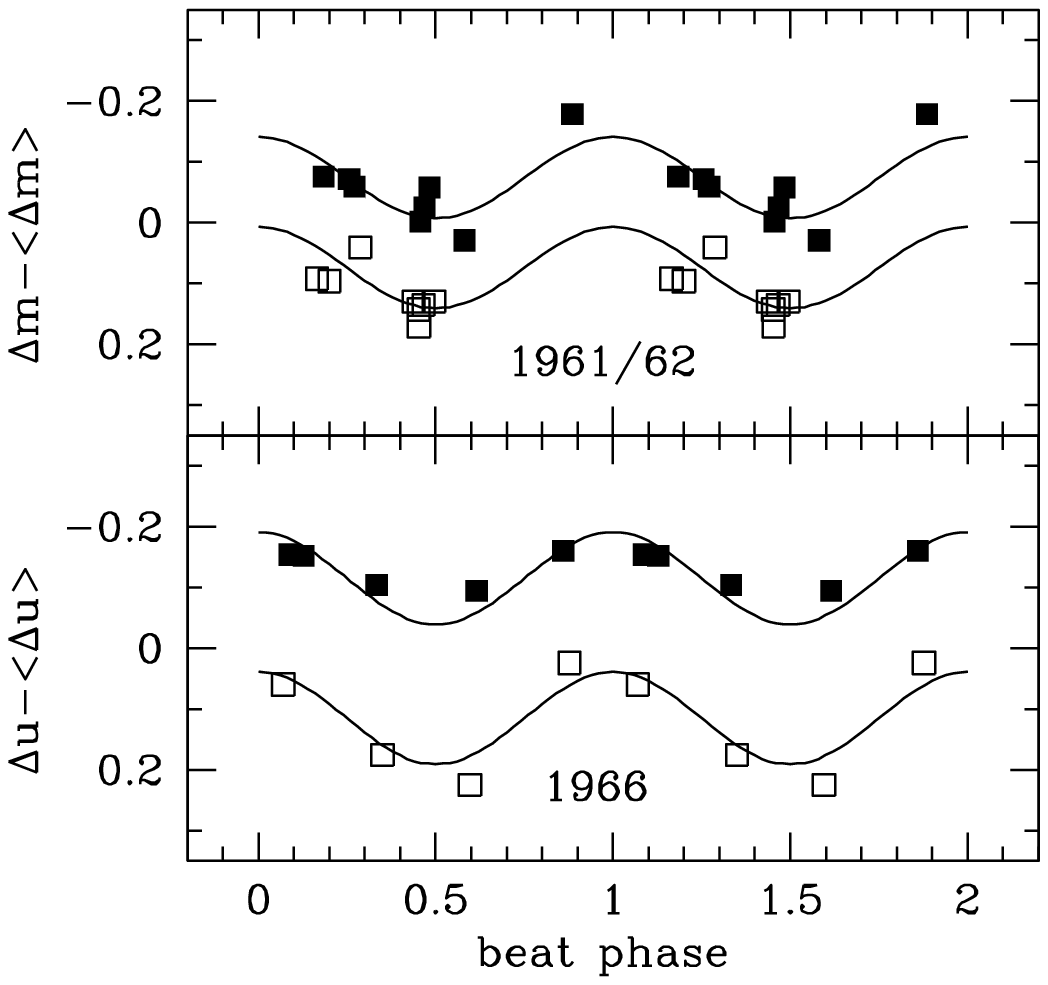} 
\vskip -5.5truecm
\FigCap { Maximum ({\it filled squares}) and minimum ({\it open squares}) magnitudes 
as a function of the beat phase. 
{\it Top:} The 1961/62 Lick data. {\it Bottom:} The 1966 OHP data. 
Solid lines are the best fit cosine curves (Eq.2). }
\end{figure}

\begin{figure}[htb]
\epsfysize=10.0cm 
\hspace{1.5cm}
\epsfbox{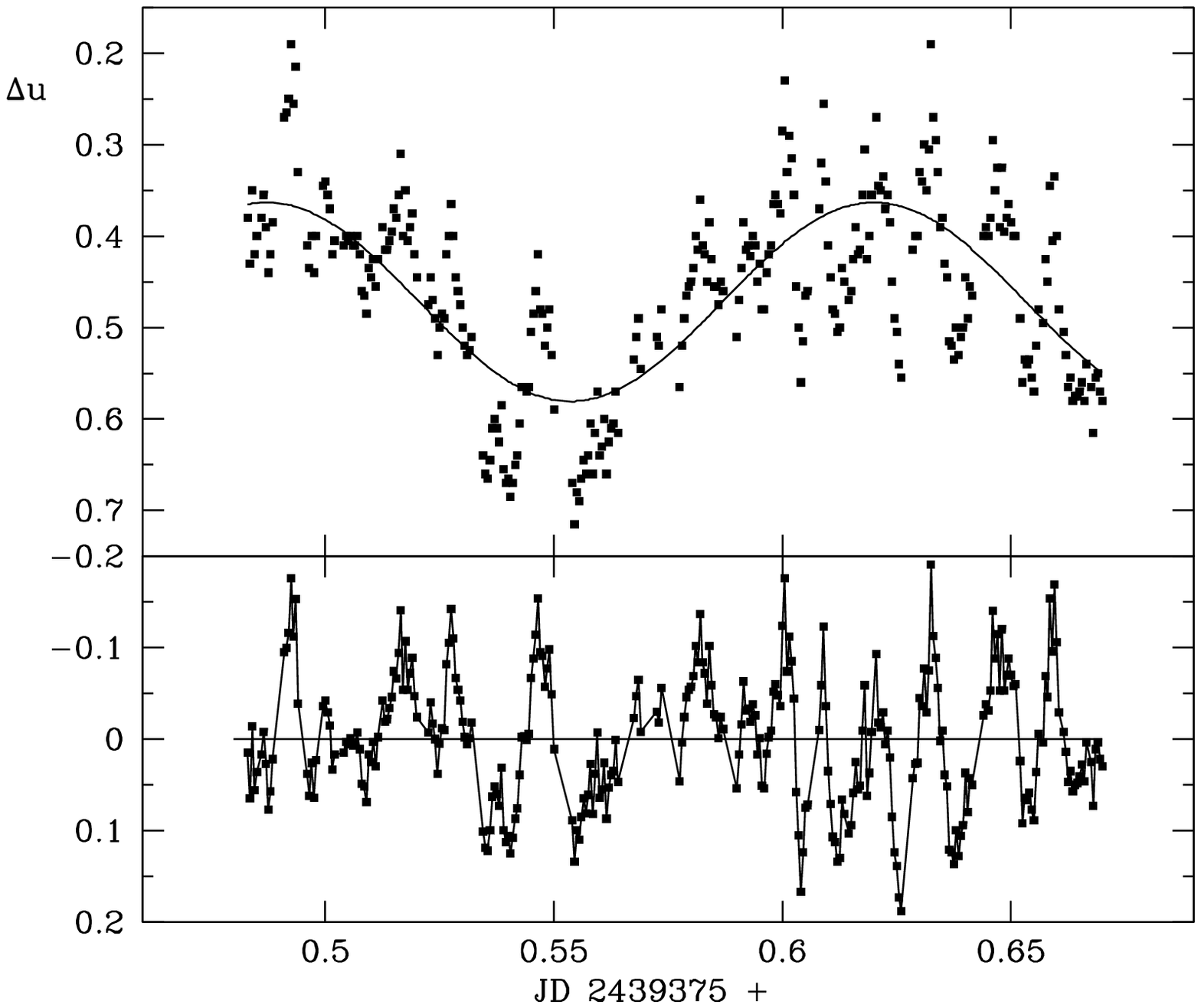} 
\vskip -3truecm
\FigCap { {\it Top:} Ultraviolet light curve of TT Ari on September 6 UT, 1966. 
Solid line is the best fit cosine curve. {\it Bottom:} The residuals after 
subtracting the cosine curve. }
\end{figure}

\section { The 1966 OHP Light Curves } 

Observations were made during five nights in August-September 1966 using the 60 cm 
reflector of the Observatoire de Haute Provence. TT Ari was observed only in ultraviolet
(defined by the Lallemand's photomultiplier Maximilien and the Corning filter C9863). 
Altogether 1280 data points were obtained. 

Fig.3 shows, as an example, the ultraviolet light curve observed on September 6 UT, 1966. Visible in the residuals after JD 2439375.60 are quasi periodic oscillations 
with $P\approx 17$ min. No oscillations, however, with $P=24$ min., suggested by 
Semeniuk et al. (1987), can be seen. 

The data were analyzed in the same way as in Section 2 giving the moments of maxima 
and minima and the corresponding magnitudes listed in Table 1. 
The moments of maxima are represented with 

\beq
{\rm Maximum}~=~{\rm JDhel.}2439360.6233(31)~+~0.132730(26)\times E~.
\eeq 

\noindent
Note that the period in 1966 was slightly shorter than in 1961/62 although this 
depends on the single maximum on JD 2439360. 

The results of the "global" fit are: $<\Delta u>=0.524\pm 0.002$, 
$A_{nSH}^U=0.115\pm 0.003$, and $A_{beat}=0.076\pm 0.003$. 
The maximum and minimum magnitudes from Table 1 and the best fit cosine curves 
obtained from the "global" fit (Eq.2) are shown in Fig.2.

\section { The "4-day" Modulation with Beat Period } 

\subsection { The Data }

A search through the literature was made for data suitable for the determination 
of amplitudes of the "4-day" modulation, resulting in the selection of the 
following sets of data. 

{\it 1987/88}. Kraicheva et al. (1997) observed TT Ari during three seasons. 
The nightly mean magnitudes $<\Delta u>$ plotted in their Fig.7 clearly 
show the "4-day" modulation. 
Regretfully, however, no information was given concerning the negative 
superhump amplitude! Fortunately, in 1987/88 TT Ari was also observed by 
Udalski (1988) and from his light curves we get $A_{nSH}=0.065\pm 0.010$. 

{\it 1988}. Tremko et al. (1996) published results of a large international 
campaign involving several observers and covering more than two months in 1988. 
They found $P_{nSH}=0.132953\pm 0.000013$d. 
Listed in their Tables 4 and 5 are moments of maxima and minima and the 
corresponding magnitudes, mostly in B. An inspection of light curves shows, 
however, that some of them were local maxima or minima unrelated to the 
superhumps and therefore had to be removed. The magnitudes $\Delta m_{max}$ and 
$\Delta m_{min}$ posed some problems. The primary comparison star used by 
Tremko et al. was BD+14{$^\circ$}336 -- the same which was used by the present 
author at Lick. However the UBV magnitude and colors obtained by them differ 
from those given in Section 2. Secondly, the values of $\Delta m_{max}$ and 
$\Delta m_{min}$ obtained by observers at Skalnate Pleso (SP) differ 
significantly from those obtained by observers at Sonneberg (SB). 
Thirdly, the values of $\Delta m_{max}$ and $\Delta m_{min}$ obtained 
observers in Kraków (KR) refer to another comparison star. Using data contained 
in Tables 4 and 5 and in Figs.6 and 7 of Tremko et al. we applied the following 
corrections: $\Delta m{\rm (SB)}=\Delta m{\rm (SP)}+0.15$, and 
$\Delta m{\rm (SB)}=\Delta m{\rm (KR)}+1.95$. 

{\it 1994}. Andronov et al. (1999) published results of another international 
campaign covering nearly three months in 1994. 
They found $P_{nSH}=0.133160\pm 0.000004$d and $A_{nSH}^B=0.0513\pm 0.0008$mag. 
Their Table 2 lists only the nightly mean magnitudes. To avoid problems with 
systematic differences between different observers we use only the results 
of a long series of observations made at the Odessa's Dushak--Eregdag Observatory. 

{\it 1996}. The long series of observations by Kraicheva et al. (1999) included 
the years 1995 and 1996 when the negative superhumps were observed and the years 
1997 and 1998 when the common superhumps were observed. 
In 1996 they found $P_{nSH}=0.13424$d -- the longest ever observed. 
Suitable for our analysis are the 1966 data: the bottom part of their Fig.4 
showing the light curves ($\Delta b$) and Fig.6 showing the nightly mean 
values; using them we determine $\Delta b_{max}$ and $\Delta b_{min}$. 

{\it 2007}. Vogt et al. (2013) presented results of continous monitoring 
of TT Ari by the MOST satellite during 10 days in 2007. They found 
$P_{nSH}=0.133103\pm 0.000036$d and $A_{nSH}^B=0.045$mag but considered 
their data insufficient for a significant detection of the "4-day" 
periodicity. In spite of that we will use the mean magnitudes plotted 
in the upper part of their Fig.2. 

\subsection { The Results }

The data described above were analyzed in the following way. In the case when 
maximum and minimum magnitudes were available they were fitted with 

\beq
\Delta m_{min}/\Delta m_{max}~=~<\Delta m>~\pm~A_{nSH}~
      -~A_{beat}~\cos (\phi_{beat}-\phi_{beat}^{max})~, 
\eeq 

\noindent 
where the $\pm$ sign refers to minimum/maximum. In the case when only nightly 
mean magnitudes were available they were fitted with 

\beq
\Delta m~=~<\Delta m>~-~A_{beat}~\cos (\phi_{beat}-\phi_{beat}^{max})~. 
\eeq 

Results are listed in Table 2 and shown in Fig.4 together with results from 
Sections 2 and 3. Listed in that Table are also the values of negative 
superhump periods $P_{nSH}$ and the corresponding beat periods $P_{beat}$ 
calculated from Eq.(3). 

\begin{figure}[htb]
\epsfysize=12.0cm 
\hspace{2.0cm}
\epsfbox{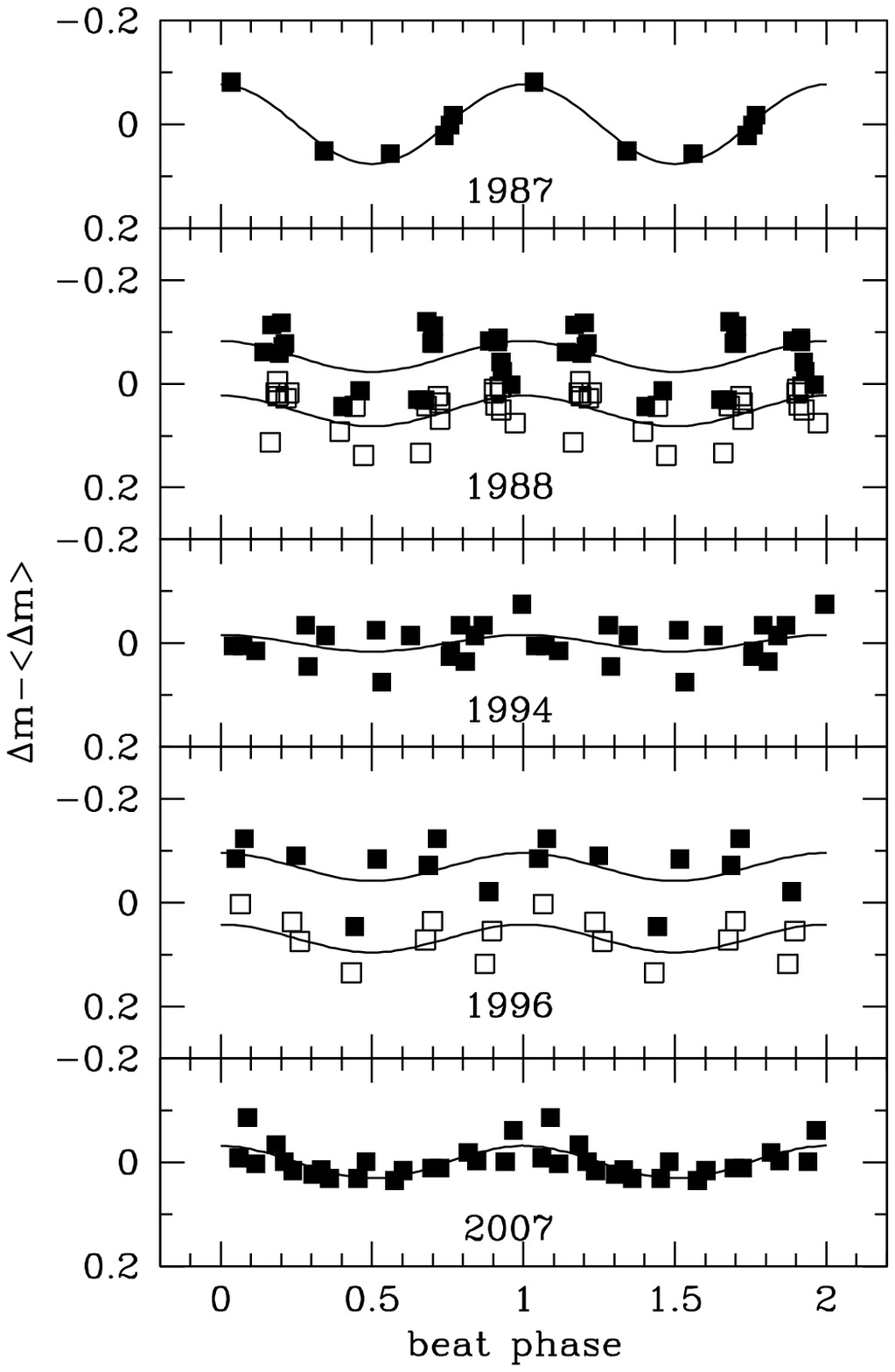} 
\vskip -2truecm
\FigCap { Maximum ({\it filled squares}) and minimum ({\it open squares}) 
magnitudes or mean magnitudes as a function of the beat phase. Solid lines are 
the best fit cosine curves (Eqs.5 and 6).}
\end{figure}

\begin{table}[h!]
{\parskip=0truept
\baselineskip=0pt {
\medskip
\centerline{Table 2}
\medskip
\centerline{ Amplitudes and Periods }
\medskip
$$\offinterlineskip \tabskip=0pt
\vbox {\halign {\strut
\vrule width 0.5truemm #&	
\enskip\hfil#\hfil\enskip&      
\vrule#&			
\enskip#\enskip&                
\vrule#&			
\enskip#\enskip&                
\vrule width 0.5truemm #&       
\enskip#\enskip&                
\vrule#&			
\enskip#\enskip&                
\vrule#&			
\enskip#\enskip&                
\vrule width 0.5 truemm # \cr	
\noalign {\hrule height 0.5truemm}
&&&&&&&&&&&&\cr
&Year&&C&&\hfil $A_{nSH}$\hfil&&\hfil $A_{beat}$\hfil&&\hfil $P_{nSH}$\hfil&&\hfil $P_{beat}$\hfil&\cr
&&&&&&&&&&&&\cr
\noalign {\hrule height 0.5truemm}
&&&&&&&&&&&&\cr
&1961/62&&V&&0.086$\pm$0.004&&0.067$\pm$0.003&&0.132896 && 3.931 &\cr
&1961/62&&U&&0.074$\pm$0.003&&0.067$\pm$0.003&&0.132896 && 3.931 &\cr
&1966   &&U&&0.115$\pm$0.003&&0.076$\pm$0.003&&0.132730 && 3.787 &\cr
&1987/88&&U&&0.065$\pm$0.010&&0.077$\pm$0.011&&0.132946 && 3.972 &\cr  
&1988   &&B&&0.052$\pm$0.007&&0.029$\pm$0.011&&0.132953 && 3.978 &\cr
&1994   &&B&&0.051$\pm$0.001&&0.016$\pm$0.014&&0.133160 && 4.172 &\cr
&1996   &&B&&0.069$\pm$0.013&&0.027$\pm$0.019&&0.134240 && 5.578 &\cr   
&2007   &&V&&0.045$\pm$0.001&&0.031$\pm$0.008&&0.133103 && 4.114 &\cr
&&&&&&&&&&&&\cr
\noalign {\hrule height 0.5truemm}
}}$$
}}
\end{table}

\section { The $A_{nSH}$ -- $A_{beat}$ and the Amplitude -- Period Correlations } 

The first, obvious conclusions based on results contained in Table 2 and shown 
in Fig.4 are:  
(1) the "4-day" modulation with beat period is always present, and  
(2) all parameters: $A_{nSH}$, $A_{beat}$, $P_{nSH}$ and -- consequently -- 
$P_{beat}$ vary significantly from one season to another. 

The two amplitudes: $A_{nSH}$ and $A_{beat}$ are compared in Fig.5 and  
it turns out that they are correlated. 
One should note, however, that indivual values of $A_{nSH}$ listed in Table 2 
correspond to three different colors. From fragmentary UBV data (Tremko et al. 
1996, Volpi et al. 1988) one finds that the amplitudes in B and V 
are practically the same, but those in U are by about 20 percent larger  
(on the other hand, however, the 1961/62 ultraviolet amplitude was {\it smaller} 
than the visual amplitude). 
Fortunately it turns out that decreasing the values of $A_{nSH}^U$ by 20 percent 
affects the correlation seen in Fig.5 only slightly. 

\begin{figure}[htb]
\epsfysize=15.0cm 
\hspace{0.0cm}
\epsfbox{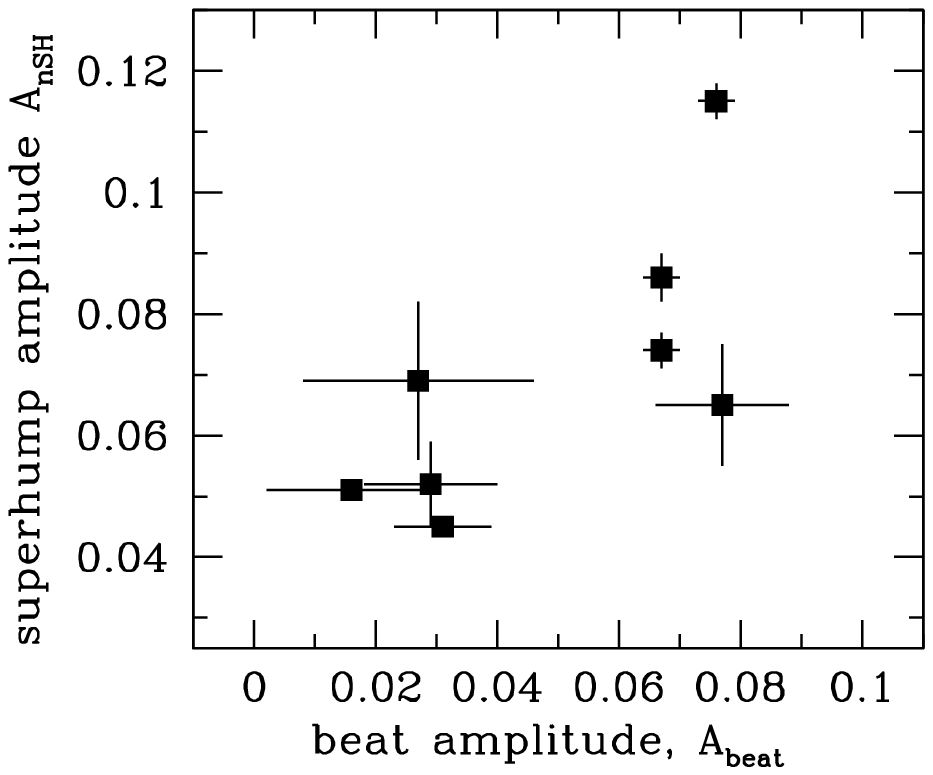} 
\vskip -8.5truecm
\FigCap { Negative superhump amplitudes $A_{nSH}$ {\it versus} 
beat amplitudes $A_{beat}$. }
\end{figure}

Shown in Fig.6 is a comparison between the amplitudes $A_{nSH}$ and $A_{beat}$ 
and the corresponding periods $P_{nSH}$ and $P_{beat}$.  
As can be seen they are also correlated. 
Worth noting is the peculiar location of the 1996 data points. 
It may suggest that the period $P_{nSH}$ and the corresponding period 
$P_{beat}$ were incorrect. Supporting this suspicion is the 
fact that $P_{nSH}=0.13424$ given by Kraicheva et al. (1999) 
differs from the mean value $<P_{nSH}>=0.132965$ based on all other 
determinations (see Table 2) by $9\sigma$, while $P_{beat}=5.578$ differs 
from $<P_{beat}>=3.992$ by $13\sigma$.  

\begin{figure}[htb]
\epsfysize=12.0cm 
\hspace{0.0cm}
\epsfbox{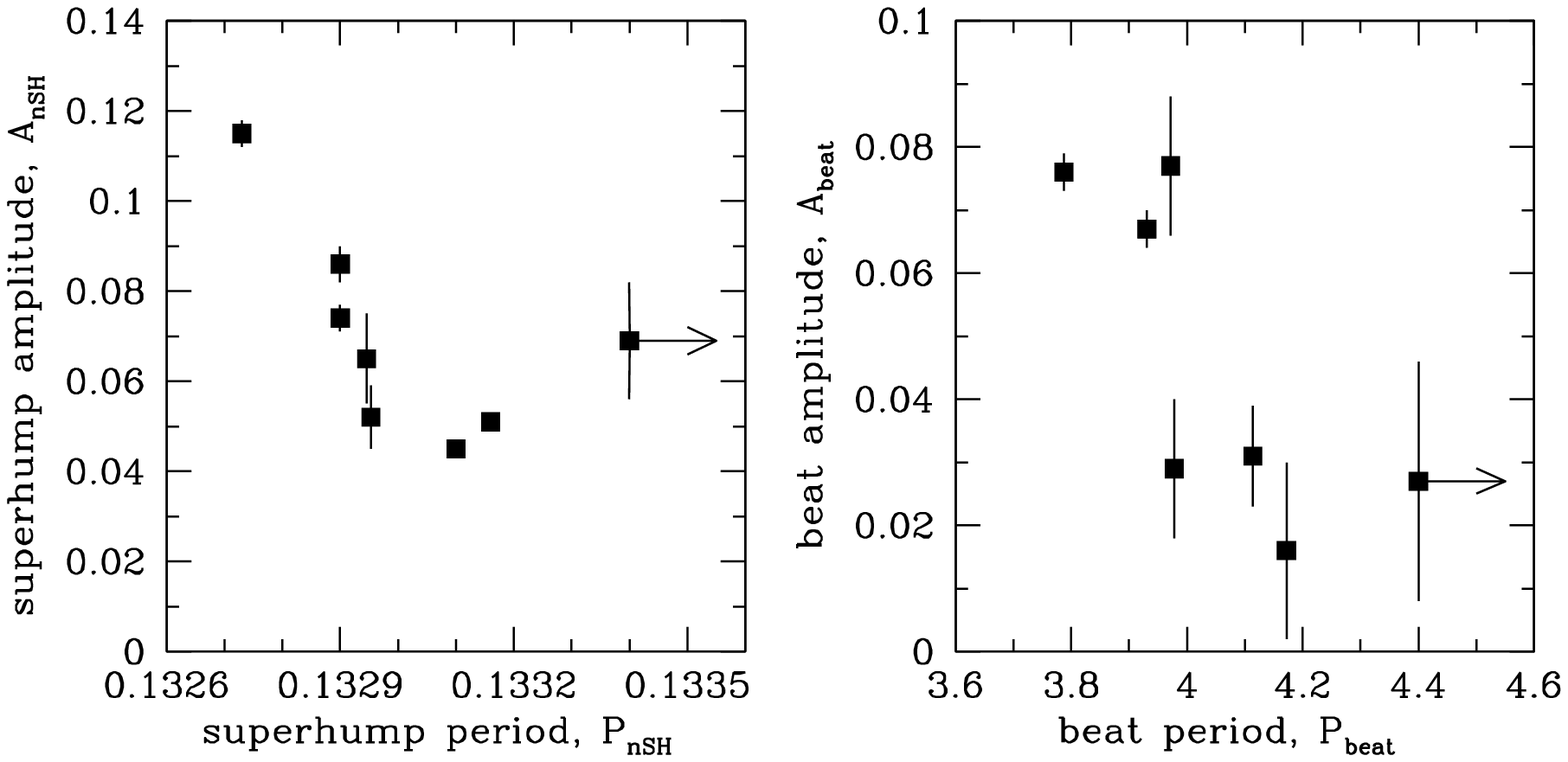} 
\vskip -6.5truecm
\FigCap { {\it left}: Negative superhump amplitudes $A_{nSH}$ {\it versus} 
periods $P_{nSH}$. {\it right}: Beat amplitudes $A_{beat}$ {\it versus} 
periods $P_{beat}$. Arrows at symbols representing the 1966 data points 
indicate that -- with $P_{nSH}=0.13424$ and $P_{beat}=5.578$ -- they should 
actually be located far to the right, outside the plot limits. } 
\end{figure}

It can be hoped that the existence of those correlations will be confirmed 
by results of two series of observations of TT Ari in 2012: with the MOST 
satellite and by the AAVSO observers (see Vogt et al. 2013). Using values 
of the negative superhump period $P_{nSH}=0.132883$ given by Vogt et al., 
and the corresponding $P_{beat}=3.916$, we predict (see Fig.6) that the 
two amplitudes, $A_{nSH}$ and $A_{beat}$, should be fairly large.

\section { Discussion } 
 
According to the commonly accepted interpretation of negative superhumps 
(Montgomery 2009ab, Wood, Thomas and Simpson 2009, and references therein) 
they are due to modulated dissipation of the kinetic energy of the stream 
as it collides with the surface of the tilted precessing disk. 
This "tilted-disk model" predicts that the negative superhump amplitude 
$A_{nSH}$ should depend on disk tilt (Montgomery 2009a). 
Furthermore, as the inclination of the disk with respect to the observer 
changes with the precession period its observed luminosity is expected to be 
modulated with $P_{prec}$. 

The beat amplitude observed in TT Ari (see Table 2 and Fig.5) varies between 
$A_{beat}=0.02$ and 0.08 mag. Using Eq.(28) from Smak (2009) with $i=29^\circ$ 
(Wu et al. 2002) we find that this corresponds to variations of the tilt angle 
between $\delta =1^\circ$ and $5^\circ$. 

The correlation between $A_{beat}$ and $A_{nSH}$ (Fig.5) provides further 
support for the "tilted-disk model". The significance of other correlations, 
however, between the amplitudes and periods (Fig.6), is not clear.

\begin {references} 


\refitem {Andronov, I.L., Arai, K., Chinarova, L.L., Dorokhov, N.I., Dorokhova, T.N., 
      Dumitrescu, A., Nogami, D., Kolesnikov, S.V., Lepardo, A., Mason, P.A., 
      Matsumoto, K., Oprescu, G., Pajdosz, G., Passuelo, R., Patkos, L., Senio, D.S., 
      Sostero, G., Suleimanov, V.F., Tremko, J., Zhukov, G.V., Zo{\l}a, S.} 
      {1999} {\AJ} {117} {574} 

\refitem {Andronov, I.L., Burwitz, V., Chinarova, L.L., Gazeas, K., Kim, Y., 
      Niarchos, P.G., Ostrova, N.I., Patk{\'o}s, L., Yoon, J.N.} {2005} 
      {\it Inf.Bull.Var.Stars} {~} {5664} 

\refitem {Cowley, A.P., Crampton, D., Hutchings, J.B., Marlborough, J.M.} 
      {1975} {\ApJ} {195} {413} 

\refitem {Herbig, G.H.} {1961} {\it Private information} {~} {~}

\refitem {Hudec, R., Huth, H., Fuhrmann, B.} {1984} {Observatory} {104} {No.1058, 1}

\refitem {Kim, Y., Andronov, I.L., Cha, S.L., Chinarova, L.L., Yoon, J.N.} 
     {2009} {\AA} {496} {765}

\refitem {Kraicheva, Z., Stanishev, V., Iliev, L., Antonov, A., Genkov, V.} 
      {1997} {\AA Suppl.Ser.} {122} {123} 

\refitem {Kraicheva, Z., Stanishev, V., Genkov, V., Iliev, L.} 
      {1999} {\AA} {351} {607}

\refitem {Montgomery, M.M.} {2009a} {\MNRAS} {394} {1897}

\refitem {Montgomery, M.M.} {2009b} {\ApJ} {705} {603}

\refitem {Patterson, J., Thomas, G., Skillman, D.R., Diaz, M.} {1993} 
      {\ApJ Suppl.} {86} {235}

\refitem {Semeniuk, I., Schwarzenberg-Czerny, A., Duerbeck, H., Hoffman, M., 
     Smak, J., St{\c e}pie{\' n}, K., Tremko, J.} {1987} {\Acta} {37} {197}

\refitem {Shafter, A.W., Szkody, P., Liebert, J., Penning, W., Bond, H.E., 
      Grauer, A.D.} {1985} {\ApJ} {290} {707} 

\refitem {Skillman, D.R., Harvey, D.A., Patterson, J., Kemp, J., Jensen, L., 
      Fried, R.E., Garradd, G., Gunn, J., van Zyl, L., Kiyota, S., Retter, A., 
      Vanmunster, T., Warhurst, P.} {1998} {\ApJ} {503} {L67} 

\refitem {Smak, J.} {2009} {\Acta} {59} {419} 

\refitem {Smak, J., St{\c e}pie{\'n}, K.} {1969} {\it {Non-Periodic Phenomena in 
     Variable Stars}, {\rm ed. L.Detre (Budapest: Academy Press)}} {~} {p.355}

\refitem {Smak, J., St{\c e}pie{\'n}, K.} {1975} {\Acta} {25} {379} 

\refitem {Strohmeier, W., Kippenhahn, R., Geyer, E.} {1957} 
      {\it Kl.Ver{\"o}ff.Bamberg} {~} {No.18} 

\refitem {Thorstensen, J.R., Smak, J., Hessman, F.V.} {1985} {\PASP} {97} {437}

\refitem {Tremko, J., Andronov, I.L., Chinarova, L.L., Kumsiashvili, M.I., 
      Luthardt, R., Pajdosz, G., Patk{\'o}sz, L., R{\"o}{\ss}inger, S., Zo{\l}a, S.}
      {1996} {\AA} {312} {121}

\refitem {Udalski, A.} {1988} {\Acta} {38} {315}

\refitem {Vogt., N., Chen{\'e}, A.-N., Moffat, A.F.J., Matthews, J.M., Kuschnig, R., 
      Guenther, D.B., Rowe, J.F., Ruci{\'n}ski, S., Sasselov, D., Weiss, W.W.} 
      {2013} {\rm arXiv: 1307.3083} {~} {~}

\refitem {Volpi, A., Natali, G., D'Antona, F.} {1988} {\AA} {193} {87}

\refitem {Wenzel, W., Hudec, R., Schult, R., Tremko, J.} {1992} 
      {Contr.Astron.Obs.Skalnate Pleso} {22} {69}

\refitem {Williams, J.O.} {1966} {\PASP} {78} {279} 

\refitem {Wood, M.A., Thomas, D.M., Simpson, J.C.} {2009} {\MNRAS} {398} {2110}

\refitem {Wu, X., Li, Z., Ding, Y., Zhang, Z., Li, Z.} {2002} {\ApJ} {569} {418}

\end {references}

\end{document}